\documentclass{aa}  

\usepackage{graphicx}
\usepackage{txfonts}
\usepackage{color}
\usepackage{xcolor}
\usepackage{longtable}
\usepackage{txfonts}
\usepackage{rotating}
\usepackage{lscape}
\usepackage{booktabs}
\usepackage{multirow}
\usepackage{hyperref}
\hypersetup{
        colorlinks = True, %[rgb 0 105 51]
     linkcolor = {blue},
     linkbordercolor = {blue},
     anchorcolor=green,
     citecolor= {blue} }

\newcommand{\gsim}{\;\lower.6ex\hbox{$\sim$}\kern-7.75pt\raise.65ex\hbox{$>$}\;}
\newcommand{\lsim}{\;\lower.6ex\hbox{$\sim$}\kern-7.75pt\raise.65ex\hbox{$<$}\;}
\newcommand{\teff}{$T_{\rm eff}$}
\newcommand{\msun}{M$_\odot$}
\newcommand{\feh}{[Fe/H]}
\newcommand{\parsec}{\texttt{PARSEC}~}

\begin{document} 

   \title{The chemical composition of the oldest nearby open cluster Ruprecht 147.}

   \author{Angela Bragaglia\inst{1}
          \and
           Xiaoting Fu \inst{2,1}
           \and
           Alessio Mucciarelli\inst{2,1}
           \and 
           Gloria Andreuzzi\inst{3,4}
           \and
           Paolo Donati\inst{1}
          }

   \institute{INAF-Osservatorio di Astrofisica e Scienza dello Spazio, via P. Gobetti 93/3, 40129 Bologna (Italy)\\
              \email{angela.bragaglia@inaf.it}
         \and
  Department of Physics and Astronomy, Bologna University, via P. Gobetti 93/2, 40129 Bologna (Italy)
        \and             
Fundaci\'on Galileo Galilei - INAF,  Bre\~na Baja, La Palma (Spain)
        \and
        INAF-Osservatorio Astronomico di Roma - Sede di Monteporzio Catone, via di Frascati 33, 00040, Monte Porzio Catone (Italy)
}

   \date{accepted by A\&A}

  \abstract
    {\object{Ruprecht~147} (NGC\,6774) is the closest old open cluster, with a
distance of less than 300 pc and an age of about 2.5 Gyr. It is therefore well
suited for testing stellar evolution models and for obtaining precise and
detailed chemical abundance information. }
   {We combined photometric and astrometric information coming from literature
and the Gaia mission with very high-resolution optical spectra of stars in
different evolutionary stages to derive the cluster distance, age, and detailed
chemical composition. }
   {We obtained spectra of six red giants using HARPS-N at the  Telescopio
Nazionale Galileo (TNG). We also used European Southern Observatory (ESO)
archive spectra of 22 main sequence (MS) stars, observed with HARPS at the 3.6m
telescope. The very high resolution (115000) and the large wavelength coverage
(about 380-680nm) of the twin instruments permitted us to derive atmospheric
parameters, metallicity, and detailed chemical abundances of 23 species from all
nucleosynthetic channels. We employed both equivalent widths and spectrum
synthesis. We also re-derived the cluster distance and  age using Gaia
parallaxes, proper motions, and photometry in conjunction with the \parsec
stellar evolutionary models.}
   {We fully analysed those stars with radial velocity and proper
motion/parallax in agreement with the cluster mean values. We also discarded one
binary not previously recognised, and six stars near the MS turn-off because of
their high rotation velocity. Our final sample consists of 21 stars (six giants
and 15 MS stars). We measured metallicity (the cluster average [Fe/H] is +0.08,
rms=0.07) and abundances of light, $\alpha$, Fe-peak, and neutron-capture
elements. The Li abundance follows the expectations, showing a tight relation
between temperature and abundance on the MS, at variance with M67, and we did
not detect any Li-rich giant. }
{We confirm that Rup~147 is the oldest nearby open cluster. This makes it very
valuable to test detailed features of  stellar evolutionary models.}


   \maketitle
%
%________________________________________________________________

\section{Introduction}\label{intro}

\begin{figure*}
\centering
\includegraphics[width=.9\textwidth]{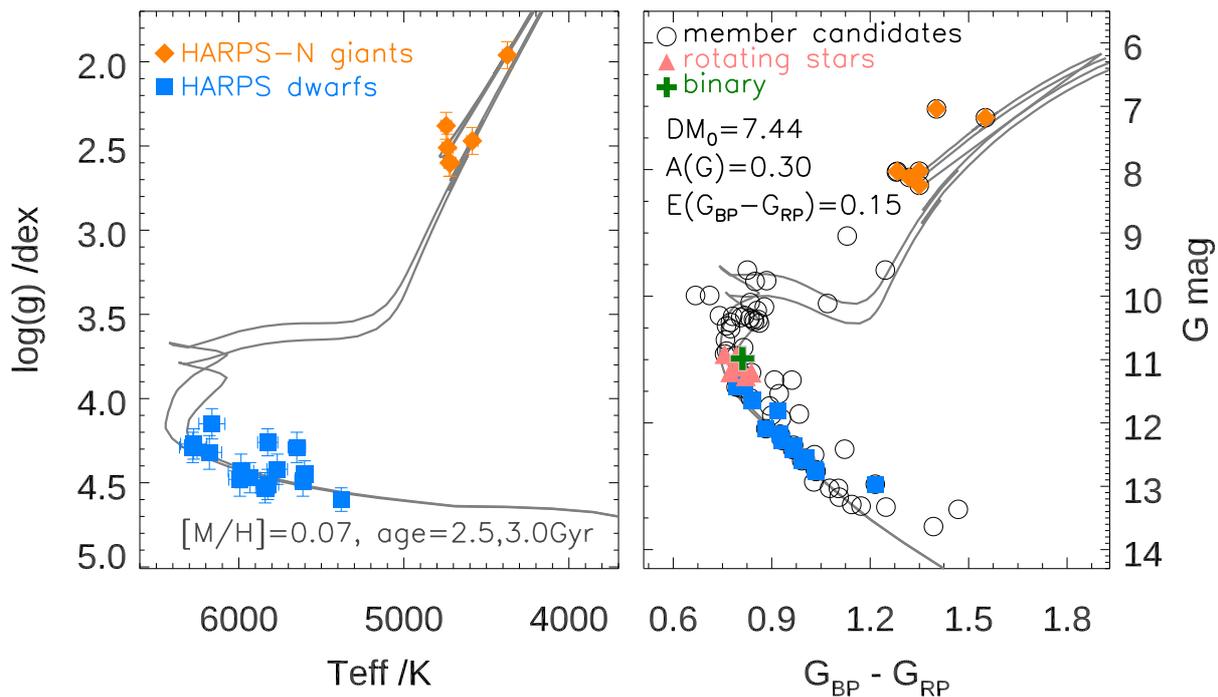}
\caption{
Left-hand panel: \teff, $\log (g)$ diagram for the 21 stars analysed spectroscopically.
Right-hand panel: CMD for Rup~147 based on Gaia DR2 photometry, with the  
stars fully analysed indicated by light blue filled squares (MS stars) and orange filled diamonds (giants). 
The stars with high rotation are indicated by pink filled triangles and the binary by a dark green plus symbol;
stars considered members by \cite{curtis13} and candidate members based on Gaia astrometry 
and without HARPS/HARPS-N spectra are shown as empty black circles.
In both panels we plot \parsec isochrones for metallicity Z=0.017 ([M/H]=0.07), 
distance 308~pc  (i.e. intrinsic distance modulus 7.44), 
and age 2.5 and 3.0\,Gyr.
A reddening $E(G_{BP}-G_{RP})=0.15$ and an extinction on the Gaia G band A(G)=0.30 are applied to the isochrones in the CMD. 
}
\label{cmd}
\end{figure*}

The Gaia mission \citep{gaiadr1a} with its legacy of astrometric and photometric
data for more than 1.3 billion objects in the Milky Way and beyond  is bringing
us what is often referred to as a revolution in Galactic astrophysics. However,
even if the Gaia Radial Velocity Spectrometer (RVS) will
provide radial velocity for a few million stars
\citep[e.g.][]{katz18,marchetti18} and chemical abundances for the brightest
among them, the spectroscopic capabilities of Gaia are limited. This leaves
space for complementary projects from the ground, such as for example the large
spectroscopic surveys Gaia-ESO \citep{gilmore12,randich13}, GALAH
\citep{martell17} and others both on-going and future, and to smaller programmes
concentrating on high precision velocities and detailed chemical composition.
The synergy with Gaia will enhance results for all Galactic populations, in
particular for the stellar clusters within the few kiloparsecs where the Gaia
astrometry reaches the highest precision and high-resolution spectra of good
quality are obtainable. This is the case for many open clusters (OCs) which can
then  be used to test the stellar evolutionary models on which, ultimately, age
determinations are based; see for example a first application combining Gaia and
Gaia-ESO in \cite{randich18}. Furthermore, detailed abundances of elements of
all nucleosynthetic chains in different evolutionary phases are important to
test the ``subtleties'' of stellar models, such as diffusion and mixing (e.g.
\citealt{onehag14,smiljanic16,bertelli_motta}). \object{Ruprecht~147}, an old
and very close OC, represents an ideal case for these studies.

Ruprecht~147 received very little attention until recently, despite being
recognised as an old (age about 2.5 - 3 Gyr) and very close (175-300 pc)
cluster  in the \cite{daml02} and \cite{kharchenko05,kharchenko13} catalogues.
High-resolution spectra of three giant stars were obtained by \cite{pakhomov},
who derived atmospheric parameters, a metallicity slightly above solar
([Fe/H]=$0.11\pm0.07$ averaging the three stars), a mean radial velocity (RV) of
$42.5\pm3$ km~s$^{-1}$, and abundances of many elements (light, $\alpha$,
Fe-peak, and n-capture). Spectra of eight giant stars were obtained by
\cite{carlberg14} to measure radial and rotation velocities; she derived a mean
RV=$42.5 \pm 1.0$ km~s$^{-1}$ and found that five of the targets are good
candidate members, based on their RV.  Two of the stars have also been studied
by \cite{pakhomov} and four are in common with our sample; comparison of results
will be presented {\bf below}. Two stars in Rup~147 were studied by
\cite{brewer16} among about 1600 F, G, and K stars observed in a search for
planets. Spectra were obtained with HIRES@Keck and analysed using spectral
synthesis, determining atmospheric parameters, projected rotational velocity,
and abundances for 15 elements. The stars are CWW 21 = SPOCS 3038 and CWW 22 =
SPOCS 3049, where SPOCS is the identification in \cite{brewer16}, and they are
not among our targets. They are solar-type stars, with [Fe/H]=+0.23.

The most relevant paper on this cluster is by \cite{curtis13}. The authors,
identifying its possible role as a ``benchmark'' cluster, given its proximity
and old age,  presented a comprehensive study of Rup~147, combining photometry,
high-resolution spectroscopy, and literature astrometry. \cite{curtis13}
selected possible astrometric members and conducted an RV survey  with three
different high-resolution spectrographs, finding about 100 candidate members and
about 10 binaries or suspected binaries. The average RV for Rup~147 is 41.1
km~s$^{-1}$. They also collected higher-signal-to-noise(S/N) spectra of six
stars,  mostly in the main sequence (MS) evolutionary phase, which could be used
for chemical analysis. On the basis of three of these stars, they derived an
average [Fe/H]$=0.07\pm0.03$, in good agreement with the \cite{pakhomov}
result.  Using deep MegaCam@CFHT photometry and different sets of stellar models
they also studied the cluster colour-magnitude diagram (CMD) and deduced an age
of about 2.5-3.0 Gyr and a distance of about 300 pc. We used information on
membership based on \cite{curtis13} to select our targets (see following
section).

The first Gaia data release (Gaia DR1, \citealt{gaiadr1b}) 
contained also  the Tycho-Gaia astrometric solution (TGAS), that is, a subset of  bright stars for which proper motions (PM) and parallaxes ($\varpi$) are derived using Hipparcos and Tycho-2 positions as first epoch. While Rup~147 is not among the 19 validation OCs studied by \cite{gaia_oc},  data for many stars towards its position were available in TGAS.  \cite{cantat18a} tried to characterise the open clusters in the solar neighbourhood (within 2 kpc) using TGAS parallaxes  and PMs complemented by UCAC4 PMs and 2MASS photometry \citep{ucac4,2mass}. They found 63 astrometric members within a radius of 3 deg and determined the following average values: $\varpi=3.26\pm0.09$ mas, $\mu_\alpha=-1.04\pm0.18$, and $\mu_\delta=-26.85\pm0.20$ mas~yr$^{-1}$ (PMs come from UCAC4). They also determined age from isochrones
for about one fifth of their 129 clusters, but unfortunately Rup~147 is not among them.  \cite{yen18}, combining information from all-sky ground-based photometry, TGAS, and HSOY \citep{hsoy}, derived fundamental parameters for 24 nearby OCs.
For Rup~147 they found: $\varpi=3.53\pm0.23$ mas (i.e. distance 265 pc), $\mu_\alpha=-1.48\pm0.26$, $\mu_\delta=-26.92\pm0.18$ mas~yr$^{-1}$, E($B-V$)=0.059, and an age of 725 Myr. They adopted \parsec isochrones with solar metallicity \citep{parsec} for their analysis; their values for reddening and especially age are not consistent with literature values or with the findings in the present paper. They acknowledge the discrepancy, but do not give a fully convincing explanation. In fact, their procedure initially found a young age for the cluster (the one published), due to the inclusion of blue stragglers in the fit, which they try to manually exclude. However, they might have  excluded too many stars close to the turn-off, resulting in an old age (about 6 Gyr).
Ruprecht~147 is present in the second Gaia data release (Gaia DR2) \citep{gaiadr2,cantat18b} and we used those
$\varpi$ and PM values in the present paper.
Parameters based on Gaia DR2 astrometry and photometry were derived by one of the validation papers \citep[][see their Table 2, where the cluster is indicated by the alternate name of NGC~6774]{babusiaux} using PARSEC isochrones and literature metallicity. They found distance modulus=7.455, log(age)=9.3, and $E(B-V)$=0.08, based on 154 candidate members. These values compare well with our findings (see Sect.~\ref{cluster_param}); the age and reddening are slightly smaller, while the adopted metallicity, [Fe/H]=0.16, is higher.

The paper is organised as follows: Section 2 describes the data, both proprietary and archival; Section 3 concerns cluster parameters derived using photometry and astrometry from Gaia; Section 4 deals with atmospheric parameters and chemical abundances; Section 5 presents a comparison with literature results; Section 6 discusses some elements in more details; and finally Sect. 7 summarises and puts our results in the context of our current understanding. 
   
%--------------------------------------------------------------------

\begin{figure}
\includegraphics[width=0.5\textwidth]{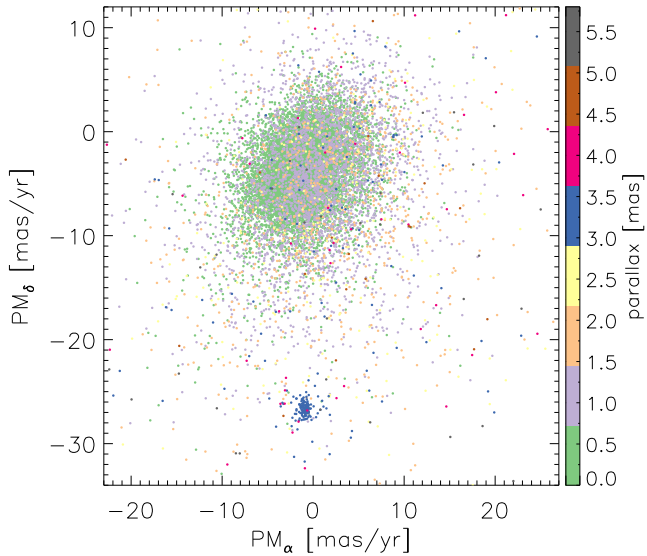}
\caption{Proper motion distribution for stars in a region of radius 80 arcmin around the nominal centre of Rup~147, 
coloured according to parallax \citep{gaiadr2}. 
Candidate member stars in Rup~147 show a well clustered distribution roughly centred on $PM_{\alpha},PM_{\delta} \approx -1,-26$ and $\varpi \approx3$. }
\label{astrom}
\end{figure}

\begin{figure*}
\centering
\includegraphics[scale=0.85]{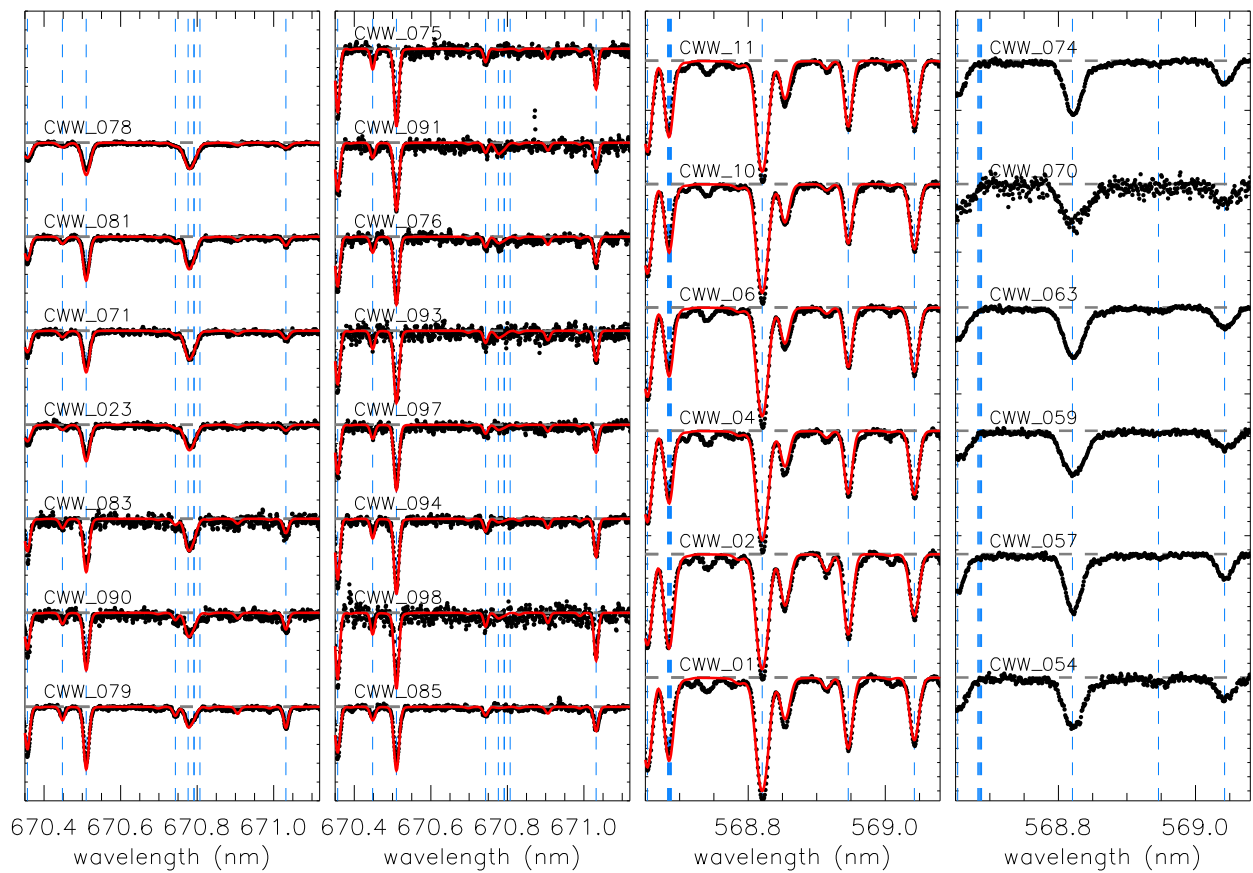}
\caption{ A small spectral region close to the Li~{\sc i} 607.8nm line for our 15 MS sample stars (first and second panels), and the spectral region close to the Na~{\sc i} 568.8nm line for the 6 giant stars observed with HARPS-N (third panel). The vertical lines indicate all the Y/U lines we used in the analysis.
 We also show examples of synthetic spectrum fits for the 21 stars analysed (see text), while the 6 stars close to the MSTO excluded from analysis are shown separately (fourth panel).}
\label{spec}
\end{figure*}

\section{The data}\label{data}

We gathered spectra of six evolved stars of Rup~147, selected among the most probable single members according to \cite{curtis13}. A log of the observations and of basic parameters taken from Gaia DR2 is given in Table~\ref{tab1}. 

We used the very-high-resolution fibre High-Accuracy
Radial velocity Planet Searcher in North hemisphere (HARPS-N) spectrograph, mounted at the Telescopio Nazionale Galileo (TNG) in La Palma, Canary Islands (programme A33DDT0). HARPS-N covers the spectral range 383-693 nm, with resolution R=115000.
The spectra were reduced automatically using the Data Reduction Software (DRS) which supplies science-quality data. The basic processing steps comprise bias subtraction, spectrum extraction, flat fielding, and wavelength calibration. The spectra were corrected for barycentric motion. 

Furthermore, we downloaded the ESO archive spectra of  22 MS stars obtained with HARPS at the ESO 3.6m telescope to search for Neptune-size planets (original programmes 091.C-0471, 093.C-0540,  and 095.C-0947). Also in this case, stars were selected among good candidate members from \cite{curtis13}. These spectra are generally of low S/N (average value about 20) but there are several/many exposures for each star (from 2 to 58 individual spectra, with a mean of 15). Information on the archive stars based on Gaia DR2 is found in Table~\ref{tab2}. We downloaded the Advanced Data Products  (ADP) spectra. The HARPS echelle data were reduced automatically using the DRS pipeline developed by the HARPS consortium, corrected for barycentric motion, and sky subtracted. The spectral coverage is essentially the same as HARPS-N, that is, 378-691 nm, and the resolution is also R=115000.  The HARPS spectra were combined to enhance the S/N (see Table~\ref{param}) and the chemical analysis was done on the combined spectra.

We measured the RV on the individual spectra using iSpec \citep{ispec} and a line list; results and errors are given in Table~\ref{param}, where we show the average values for the MS stars. We also obtained v$\sin i$, again using iSpec; most of the stars are slow rotators (see Table~\ref{param}). However, as shown in
the right panel of Fig.~\ref{cmd}, we eliminated the six stars closer to the MS turn-off (MSTO), because their rotation velocity makes their lines wider and more subject to blends. 

The MS stars show generally a constant RV, however, we found two interesting cases among them: a) CWW~58 is clearly a binary, with an RV variation $>5$ km~s$^{-1}$ in the 9 spectra obtained over a two-year interval; and b) CWW~71 shows a linear trend in its RV, which changes from 42.17 to 41.23  km~s$^{-1}$ for the 11 spectra, again obtained over approximately 2 years.  Neither one was indicated as problematic in \cite{curtis13}. We excluded star CWW~58 from further analysis but retained star 71. We have only one spectrum for the giants, so we cannot state that they are single stars; comparison with literature values cannot be conclusive because they are based on spectra of lower resolution and precision. Furthermore, systematics between the different analyses could hide small differences such as the ones we found for the two stars discussed above.

\section{Cluster parameters }\label{cluster_param}

The stars observed are shown in Fig.~\ref{cmd} (right panel) in the CMD based on Gaia G band, $BP$ and $RP$ data.
The targets were selected among high-probability members, 
based on RV and ground based proper motions  \citep[see][for details]{curtis13} and confirmed 
as members by Gaia DR2 PM, $\varpi$ values  a posteriori. The stars observed define the cluster sequence very well.

For this further assessment of their membership we downloaded $\varpi$ and PM values for 
stars in a region of 80 arcmin in  radius  around the cluster centre. 
In Fig.~\ref{astrom} we show the proper motions for stars with G\,<15 mag, 
colour-coded using the parallax; 
Rup~147 is well isolated from field stars. 
All  stars in our spectroscopic sample are also included in Gaia DR2 
(see Tables~\ref{tab1}, \ref{tab2}) and their mean $\varpi$ (3.25$\pm0.09$ mas) 
and PMs ($PM_{\alpha}=-0.95\pm 0.78$ mas/yr, $PM_{\delta}=-26.53\pm0.65$ mas/yr)
are in very good agreement with the cluster averages based on TGAS or DR2 \citep{cantat18a,cantat18b} .
All other stars that satisfy the Rup~147 parallax and PM are considered as member candidates (black circles in Fig. ~\ref{cmd}). 

The averages were computed as simple mean values, 
without taking into account the correlations between astrometric parameters, which do not have a relevant impact. 
Furthermore, we used the derived distance mainly to find good starting points for \teff \, and gravity for the spectroscopic analysis, 
in combination with the \parsec isochrones 
\citep[version 1.2S from CMD 3.0 web input form \footnote{http://stev.oapd.inaf.it/cgi-bin/cmd\_3.0},][]{parsec, chen2014}. 

From the isochrone fit in Fig.~\ref{cmd} we estimate that Rup~147 is 2.5 to 3.0
Gyr old,  adopting distance=308 pc from Gaia DR2, which translates to
$(m-M)_0$=7.44,  metallicity Z=0.017 ([M/H]=0.07, see \citealt{curtis13}), 
which is in good agreement with what we find (see Sect.~\ref{chem}), extinction
$E(G_{BP}-G_{RP})=0.15$, and absorption in the Gaia G band A(G)=0.30.  By
assuming a standard extinction law \citep[R$_V$=3.1,][]{cardelli89} and taking
$A(G)= 0.85926 \,A(V),~ A(G_{BP})= 1.06794\,A(V),~ A(G_{RP})= 0.65199 \,A (V)$
~from the \parsec website,   the reddening in (B-V) colour is
$E(B-V)=0.775\,E(G_{BP}-G_{RP})=0.113$, and the extinction in V band is
$A_V=0.35$,  which is in between the values of 0.46 and 0.25 from 
\cite{pakhomov} and \cite{curtis13},
respectively.  Further refinements are not required for the main goal of this
paper, which focuses on the detailed chemical properties.

Adopting the Gaia DR2 parallax of the member candidates, we calculate the
heliocentric Galactic coordinates of Rup~147: $X=280.21\pm8.14$\,pc (towards the
Galactic centre), $Y=106.90\pm4.23$\,pc (towards the local direction of rotation
in the Galactic plane), and $Z=-70.00\pm2.38$\,pc (towards the north Galactic
pole). The Galactic radius of this cluster is $R_{GC}=8.28$ kpc. Its iron
abundance (\feh=0.08) is in good agreement with the expectations at its
Galactocentric radius \citep[see e.g. the homogeneous samples
in][]{donati15,netopil16,reddy16}.

We then confirm once more that Rup~147 is the only old and nearby OC;  next OC
close-by and older than 1 Gyr is  NGC~752 (age and distance about 1.6 Gyr and
450 pc, respectively)  and we need to reach approximately 900 pc to find an OC
older than Rup~147, that is M67.  Rup~147 is therefore very important as a
benchmark cluster, as remarked by \cite{curtis13},  and efforts to determine its
detailed properties through photometry, astrometry, high-resolution
spectroscopy, and modelling are welcome.

\section{Atmospheric parameters and chemical abundances}
\label{chem}

To derive the atmospheric parameters we used the equivalent widths (EWs) of iron
lines, both neutral and ionised,  employing MOOG \citep{moog} via iSpec.  Our
analysis was done assuming local thermodynamic equilibrium (LTE) and using the
MARCS model atmospheres \citep{marcs}.  We used the public Gaia-ESO line list
\citep[][]{heiter15,ges_linelist}, which is based on VALD3 data \citep{vald}, 
selecting only the Y/Y lines, that is, the most isolated ones with the most
robust atomic data.  We followed the classical spectroscopic method to derive
temperature \teff, gravity $\log g$, microturbulent velocity $\xi$, and the iron
abundance [Fe/H].  \teff \, is obtained eliminating trends between the line
abundances and the excitation potentials (excitation equilibrium),  $\log (g)$
requiring that Fe {\sc i} and Fe {\sc ii} give the same abundance (ionisation
equilibrium),  and $\xi$ was obtained by minimising the slope of the relation
between line abundances and EWs.  The stellar parameters are given in
Table~\ref{param}, together with the uncertainties, based on the uncertainties
in the slopes of the three relations. With the \teff \, and $\log (g)$ values,
we derive stellar mass from isochrones for our sample stars,  the results are
also listed in Table ~\ref{param}.

We obtained an average [Fe/H]=0.08 (rms 0.07) dex for Rup~147.  If we divide the
giants from the dwarfs to take into account possible (small) effects of
diffusion \citep[see e.g.][both on M67]{onehag14,bertelli_motta}, we have
[Fe/H]=0.10 (rms 0.06) and 0.07 (rms 0.08) dex for the six giants and the 15 MS
stars, respectively. 

We derived the abundances of 23 species, including Li, light, $\alpha$, Fe-peak,
and neutron capture elements.  We employed iSpec, again using the MOOG choice
and the GES public line list, choosing Y/U lines. Given the (much) smaller
number of lines available, we relaxed the criterium adopted for iron and also
used lines for which blending had not been checked by the GES consortium;
however, our spectra have a larger resolution and we inspected dubious cases. We
employed spectrum synthesis for all lines, including hyper-fine structure (HFS).
In Fig.~\ref{spec} we show examples of the region near the Li {\sc i} line for
the 15 MS stars and near Na {\sc i} for the 6 giants and the 6 MS stars excluded
from further analysis because of their larger rotation velocity. We checked that
the line list and the synthesis reproduced the solar abundances  using the
spectrum ``HARPS.Archive\_Sun-4'' from the library of the Gaia FGK benchmark
stars  \footnote{https://www.blancocuaresma.com/s/benchmarkstars}
\citep{benchmark}.  We found a difference for only three elements (Cu, Ba, and
Eu),  so we corrected the cluster abundances by these offsets based on our
derived solar abundance. Finally, we visually checked a few lines in case of
large dispersion in the line-by-line abundances. 

All abundances were obtained using LTE and are reported in Tables \ref{light},
\ref{heavy}, and \ref{ncap}.  Oxygen was measured from the forbidden [O {\sc i}]
630.3nm line in the six giants,  after making sure it was free from telluric
contamination.  The O triplet near 777nm is not present in the HARPS wavelength
range,  so we did not measure O in the MS stars. For Li and Na we also corrected
the LTE abundances with the prescription in \cite{lind09,lind11};  we used the
INSPECT web page\footnote{http://inspect.coolstars19.com/} deriving the non-LTE
(NLTE) corrections line by line.  In Table~\ref{light} we give both LTE and NLTE
abundances.

We derived the sensitivity to changes in stellar parameters by repeating the
analysis for one typical giant and MS star,  changing one parameter while
holding the other fixed. Results are presented in Table~\ref{sens}. 

\begin{figure*}
\centering
\includegraphics[scale=0.96]{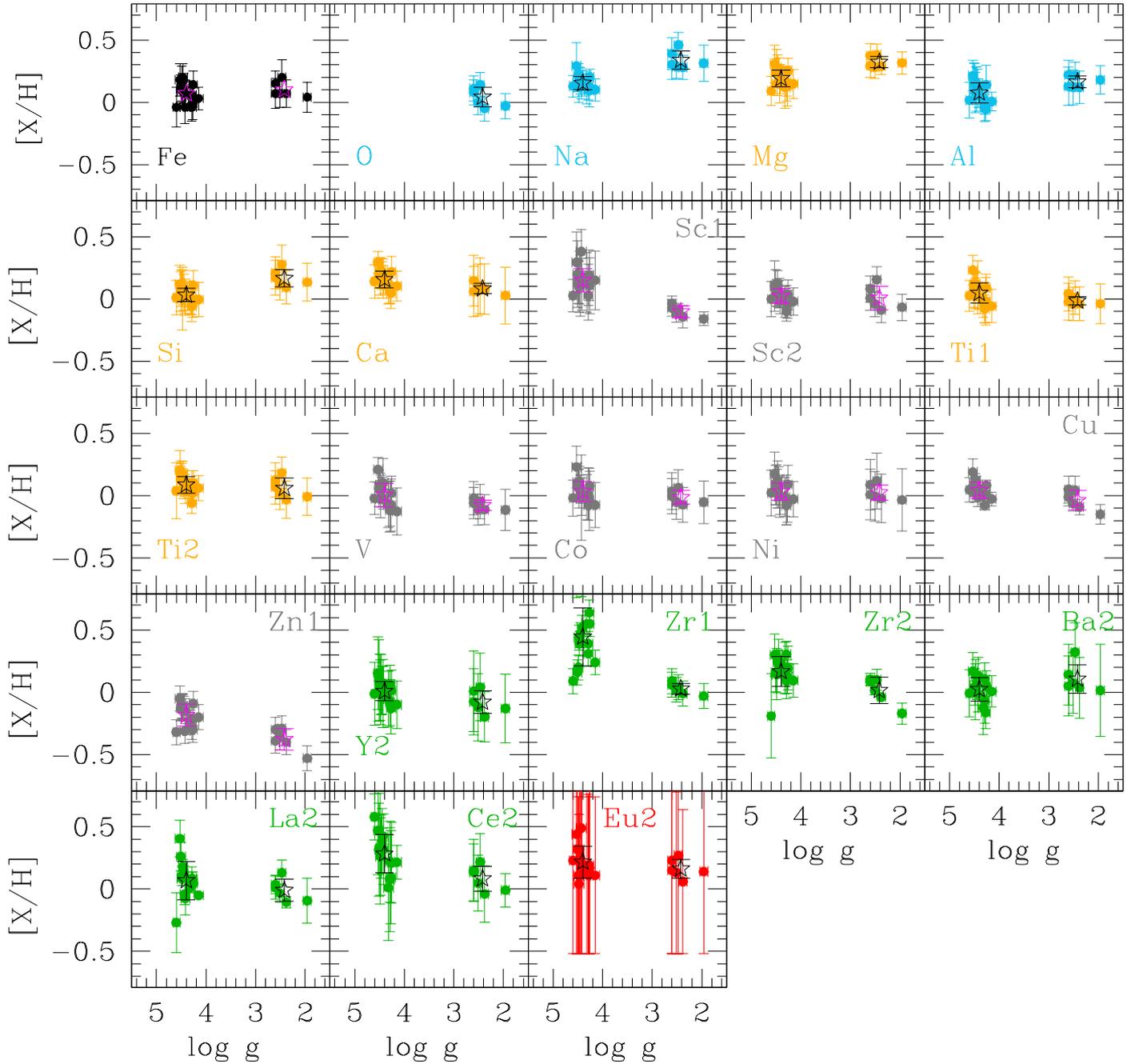}
\caption{We show here the run of the derived elements ([X/H]) with gravity.  Different colours indicate light, $\alpha$, Fe-peak, n-capture (slow and rapid). Starred symbols indicate the mean values for dwarfs and giants (error bars are the standard deviation).}
\label{diff}
\end{figure*}

We show in Fig.~\ref{diff} the run of [X/H] values with $\log g$ for all
elements, with the exception of Li, which will be discussed in
Sect.~\ref{discussion}.  We see that dwarfs and giants have slightly different
levels in some cases.  This is expected for Na \citep[see e.g.][ and
Sect.~\ref{discussion}]{smiljanic16,smiljanic18}  and can be explained for the
other elements by the larger uncertainties associated to the analysis of the
dwarfs (less and weaker lines)  and by the systematic differences expected from
their different atmospheres and sensitivity to details in the analysis (see e.g.
\citealt{dutra-ferreira} on the Hyades cluster).  In principle, evolutionary
differences may also be expected as a result of diffusion processes;  they have
been found for the older cluster M67 \citep[see][for Gaia-ESO, APOGEE, and GALAH
results, respectively]{bertelli_motta,souto,gao}. However, the efficiency of the
diffusion depends on the cluster age and we checked that only very small
variations are expected for an age of 2.5-3 Gyr (less than $\sim0.1$ dex in most
of the cases)  using both \parsec and \texttt{MIST} \citep{mist} stellar models.

We computed the average abundance ratios [X/Fe] for Rup~147,  given in
Table~\ref{mean},  together with the root mean square (rms), both all together
and separating dwarfs and giants.   They were obtained adopting  the reference
solar values from \cite{grevesse07}.

\begin{figure*}
\centering
\includegraphics[width=.9\textwidth]{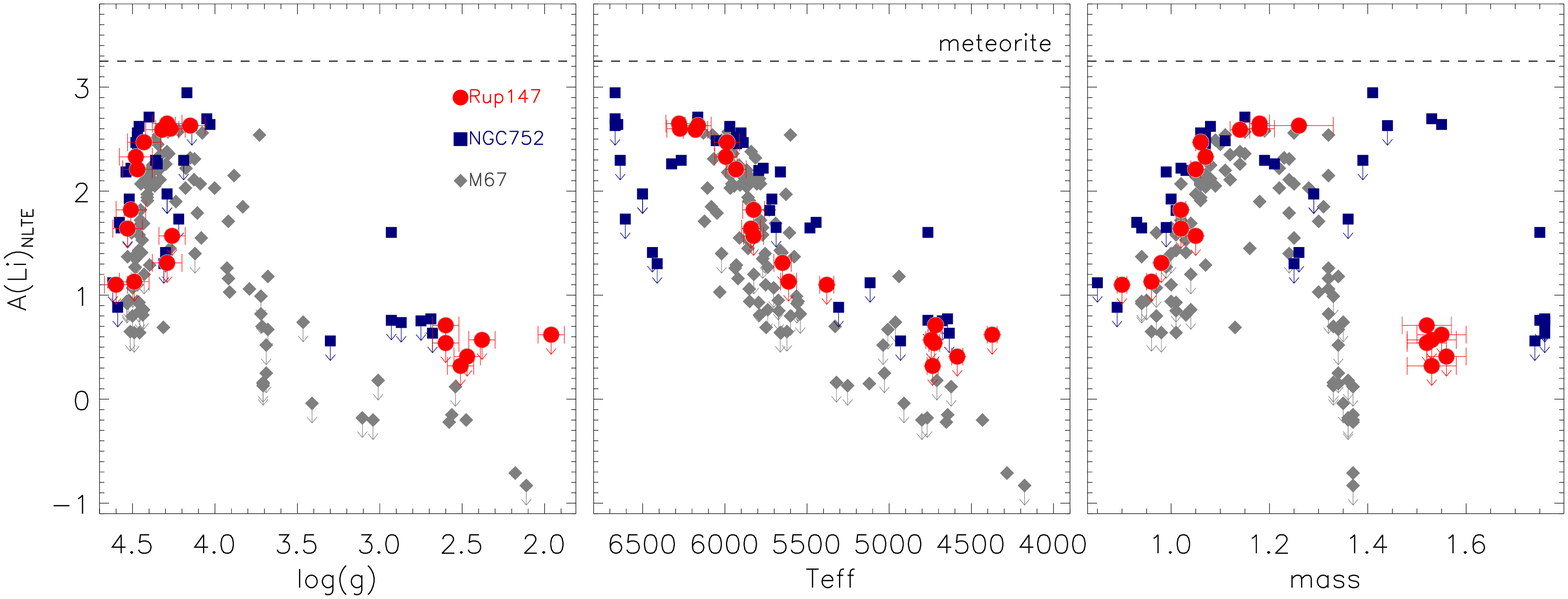}
\caption{ Li evolution as a function of log(g) (left panel), \teff (middle panel), and stellar mass (right panel).
Three clusters with similar metallicity but different ages are displayed.
The red-filled dots are our Rup~147 (2.5-3.0 Gyr) sample stars with NLTE correction, 
the blue-filled squares are NLTE-corrected NGC~752 ($\sim$ 1.6 Gyr) single stars originally from \citet{castro16}, 
and the grey-filled diamonds are Li-corrected M67 ($\sim$ 3.7 Gyr) data from \citet{pace2012}. 
The Li abundance derived from chondrite meteorites \citep[A(Li)=3.25 dex,][]{grevesse07} 
is indicated by a dashed line.
}
\label{lievo}
\end{figure*}

\section{Literature comparison}\label{lit}

We have only one star in common with \cite{pakhomov}, CWW10/HD180112,  for which
we have $4745\pm33$/$2.38\pm0.08$/$0.07\pm0.11$ for \teff, $\log g$, and [Fe/H],
compared to 4733/2.53/$0.14\pm0.06$ (see Table~\ref{conf}). Our average [Fe/H],
based only on the giants for consistency with their work, is in perfect
agreement: $0.10\pm0.06$ (six giants) compared to $0.11\pm0.07$ (three giants).

We have four stars in common with \citet{carlberg14};  the difference in RV is
within 0.5 km~s$^{-1}$ and also the $v \sin i$ values are in agreement within
the errors; see Table~\ref{conf}.

\cite{curtis13} have the most complete analysis of Rup~147 to date. We agree
with them on age (2.5 to 3 Gyr), distance (about 300 pc), and metallicity. For
the last, their average is $0.07\pm0.03$, based on five MS stars observed with
Keck/HIRES, while we have $0.07\pm0.08$ from 15 MS stars. We have two stars in
common (CWW~78, 91, see Table~\ref{conf}), their \teff \, and $\log g$ are
larger in our study than in theirs, but the metallicities are in better
agreement.

\cite{curtis18} studied star CWW 93 (hosting a sub-Neptune planet) in detail by
means of photometry and spectroscopy and also obtained spectra of a further six
solar-type stars in the cluster. All spectra were obtained with MIKE@Magellan
and were analysed using SME (Spectroscopy Made Easy, \citealt{sme}). For the
seven solar-type stars they derived [Fe/H]$=0.10\pm0.04$, while the
spectroscopically derived parameters for CWW 93 are \teff=5697~K,
$\log~g$=4.453, [Fe/H]=0.141, and $v\sin i$=1.95 km~s$^{-1}$.The mass and radius
of the star were obtained combining spectroscopic results with photometry and
the distance modulus in \cite{curtis13} and adopting three different isochrone
sets and methods.  The procedures gave consistent values and they adopted the
mean values as final choice: mass=$1.009\pm0.027$~\msun \ and
radius=$0.945\pm0.027$~R$_\odot$. For comparison,  for CWW~93 we obtain
\teff=5841$\pm57$ K, $\log g $=4.53$\pm0.09$, [Fe/H]=0.18$\pm0.13$, and $v\sin
i=1.21\pm1.29$ km~s$^{-1}$. The implied stellar mass for this star is 1.02
$\pm0.01$ M$_\odot$.

Finally, Gaia DR2 contains the RV for 18 of the 21 stars in our final list,
obtained by the Gaia RVS instrument \citep{cropper18}. The RVs are generally in
agreement, especially when the error on the  RVS measurements is small (see
Table~\ref{conf}). For all cases with a large difference, the RVS value has an
error (much) larger than 1 km~s$^{-1}$, while all our errors are one order of
magnitude smaller. The RV of the binary star CWW~58 is also similar between our
measurements and Gaia's (36.35, rms=2.21 and $35.38\pm2.06$ km~s$^{-1}$,
respectively); the Gaia pipelines did not detect this star as a possible binary.
The validation of RVs for DR2 discards stars with very high errors (20
km~s$^{-1}$) or suspect SB2 systems  \citep{katz18} and they do not apply to
star CWW~58. However, the Gaia RV is based only on two transits, so we believe
that ours is a more robust indication.

\begin{figure}
\includegraphics[scale=0.43]{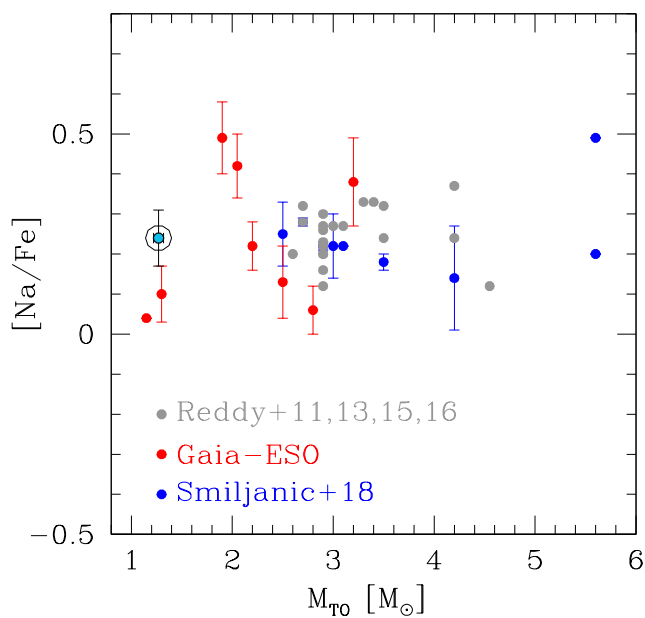}
\caption{Run of  [Na/Fe] with mass at MSTO, where each point represents a
cluster average, based only on evolved stars. Our determination for Rup~147
(average of the 6 giants) is shown as a filled light blue circle with error bars
highlighted by a larger circle. We show literature values coming from three
homogeneous groups: the Gaia-ESO survey clusters in \cite{smiljanic16},
\cite{overbeek17}, and \cite{tang17} are
shown  in red,  the young clusters in \cite{smiljanic18}
in blue, both with error bars, and clusters in \cite{reddy12,reddy13,reddy15,reddy16} in grey.}
\label{na}
\end{figure}

\begin{figure*}
\centering
\includegraphics[scale=0.96]{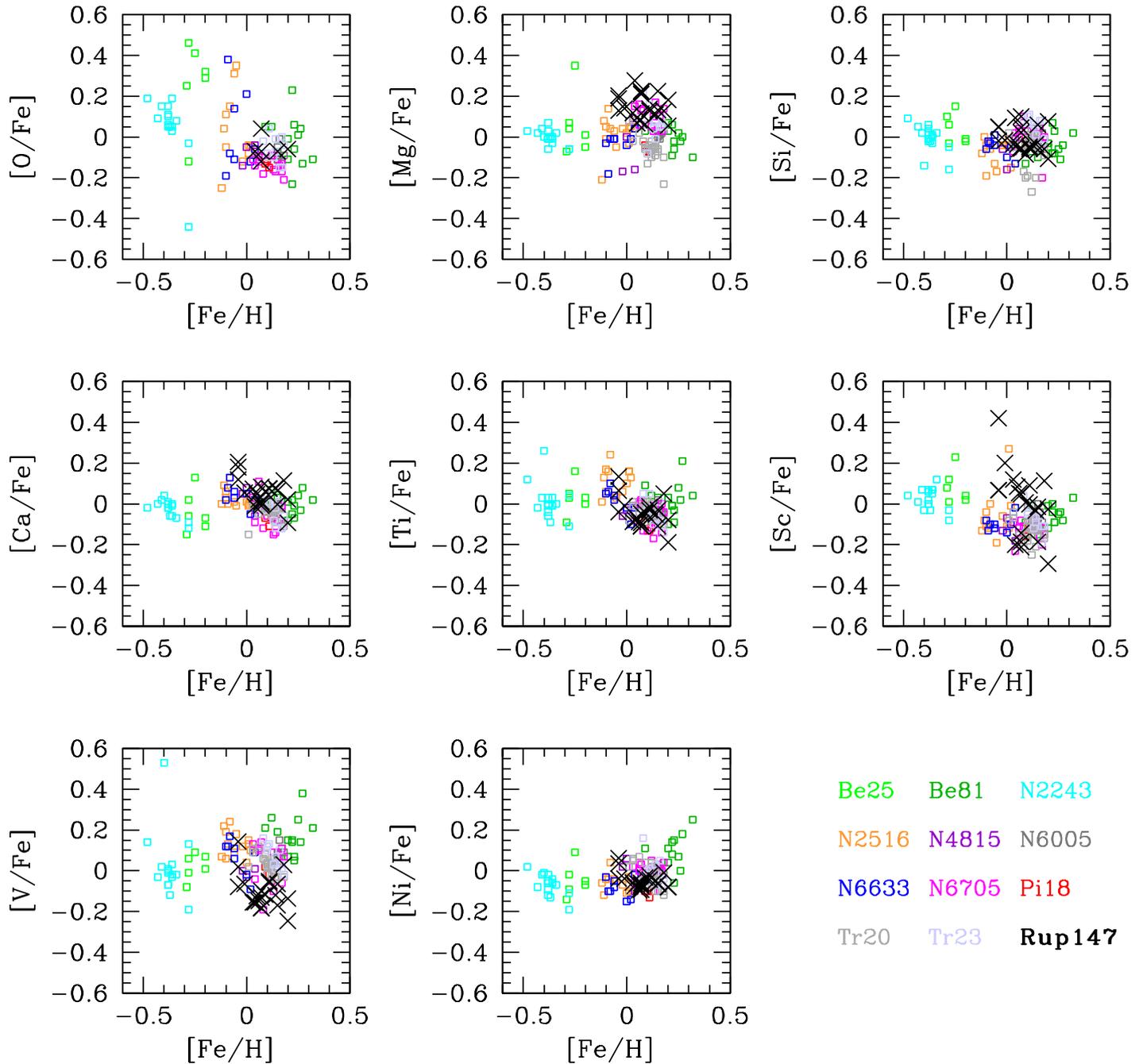}
\caption{Comparison of some of our abundance ratios (black crosses) with those
of Gaia-ESO open clusters from \cite{magrini17}, shown as open squares of
different colours (see legend). For  both samples we show the individual stars
value, not cluster averages. The agreement is good, especially once differences
in solar reference is taken into account (their solar O and Mg is higher by 0.12
dex and V lower by 0.11 dex).}
\label{ges_oc}
\end{figure*}

\section{Discussion}\label{discussion}

It has been found that Li abundance decreases as a function of age in solar twin
stars  and in open clusters \citep[see e.g.][and references
therein]{carlos16,castro16}. However, age itself does not play a key role in
this correlation, the genuine driver of the abundance dispersion is Li burning
at the bottom of the surface convection zone,  which could be indicated by
\teff. For instance, \citet{xiong2009} point out that for stars  with the same
temperature (\teff $\lessapprox$6000 K), the Li abundance decreases as age
increases. In Fig. ~\ref{lievo} we compare the Li abundance of Rup~147 to that
of two other open clusters with similar metallicity but different ages. Data for
NGC~752  is from \citet{castro16} who provide \feh=0.0\,dex and an age of
$\sim$1.6\,Gyr. Lithium abundances in \citet{castro16} are originally given in
LTE, but in Fig.~\ref{lievo} we applied NLTE corrections \citep{lind09} to them
considering a uniform microturbulence velocity $\xi$=2 km~s$^{-1}$. This does
not introduce spurious effects, since the NLTE correction for Li is not
sensitive to microturbulence; for instance,  using $\xi$=1 km~s$^{-1}$ changes
the final results by $<0.01$ dex. In the figure,  only stars marked as single
are plotted. Lithium data on M67 (\feh=0.01, age=3.7\,Gyr) from \citet{pace2012}
are already NLTE-corrected. Since several works in the literature conclude that
the Li abundance scatter in M67 may be an exception for Li evolution in open
clusters  \citep[see e.g.][]{sestito2004, xiong2009}, we put the M67 data in the
figure only for reference. The left panel of Fig. ~\ref{lievo} illustrates the
Li evolution from the MS to the giant branch for the three clusters. Though the
dwarf stars show a minor difference on Li,  the giants stars of Rup~147 and
NGC~752 have a very similar Li abundance level. Here we also notice that there
is an un-reported Li-rich giant star (H77) in NGC~752 with
A(Li)$_{NLTE}$=1.60\,dex. In the middle panel, Rup~147 shows a tight A(Li)-\teff
\, relation for the MS stars (\teff \, from $\sim$5300\,K to 6300\,K); there is
no large Li scatter as seen in M67. This tight relation supports the conclusion
of \citet{sestito2004} that M67 is the only cluster showing a large Li spread
for solar-type stars, and the Li scatter is not typical of an old open cluster.
Furthermore, all of the Rup~147 dwarf stars in our sample have
\teff\,<\,6300\,K, which is around the \teff ~border of the Li-dip at solar
metallicity (see NGC~752 in the middle panel of Fig. ~\ref{lievo}). Considering
that the turn-off \teff ~of Rup~147 is $\lessapprox$\,6400\,K (see the HR
diagram in Fig. \ref{cmd}), one can very hardly expect a dip-like pattern in the
Li-\teff~ figure of Rup\,147 even if more turn-off stars are observed in the
future. 

Compared to the MS stars of NGC~752, which is 1\,Gyr younger, Rup~147 dwarfs
present a lower Li abundance at the same temperature. The age difference on Li
abundance is also seen in the right panel of Fig. ~\ref{lievo}; with the same
stellar mass, Rup~147  MS stars (M < 1.3 \msun) have lower Li abundance compared
to NGC~752 dwarfs, a difference mostly caused by microscopic diffusion. However,
the stellar mass was derived using different stellar models for each cluster,
and this may introduce some systematic uncertainty.

The average [Na/Fe] value for MS stars is 0.08 dex, while for giants this is
0.24 dex.  This enhancement for giants is not uncommon among OCs  \citep[see
e.g.][for references]{maclean15} but is not universal \citep[e.g.][all for old
OCs]{sestito08,bragaglia12}. After excluding cases due to neglecting NLTE
effects, the enhancement may be attributed to mixing of Na to the stellar
photosphere after the first dredge-up \citep{iben67}. The amount of mixing is
then dependent on the stellar mass (and metallicity), with low-mass,
low-metallicity stars showing no changes \citep[see the observations
by][]{gratton00} and higher-mass stars showing increasing indications \citep[see
e.g. the models by][]{char_laga}. The presence and extent of Na enhancement
among OC giants has recently been studied systematically by
\cite{smiljanic16,smiljanic18}, with comparison to various stellar models. For
Rup~147, given the age, we should in principle not expect a large effect, but it
seems on the contrary to show a higher Na enhancement than two OCs of similar
age in the Gaia-ESO survey (see Fig.~\ref{na}). However, our solar Na is 6.17
\citep{grevesse07}, while \cite{smiljanic16} use 6.30; had we used the latter, 
[Na/Fe] would be at the same level as the Gaia-ESO clusters (the solar reference
iron is 7.45 for both samples). 

Apart from Li and Na, which are known to vary with evolutionary phase, how do
other elements behave in comparison to other open clusters? We compare $\alpha$
and Fe-peak elements with the results of 11 OCs homogeneously analysed by the
Gaia-ESO survey covering almost the whole interval in metallicity of OCs and
with ages from 120 Myr to 4 Gyr, published in \cite{magrini17}.
Figure~\ref{ges_oc} shows that Rup~147 behaves well, with abundance ratios in
line with clusters of similar metallicity. Only three species look slightly
discrepant (O, Mg, and V), but the differences between Rup~147 and the Gaia-ESO
results can be explained by the different solar reference values adopted (see
figure caption).

\section{Summary and conclusions}\label{summary}

We observed six evolved stars in the old, nearby cluster Rup~147 using HARPS-N
at the TNG and retrieved spectra of 22 MS stars from the ESO HARPS archive. Our
final sample comprises the six giants  and 15 MS stars (one of the MS stars was
excluded because we found it to be a binary, and six more because they rotate
and display wide lines). We measured RVs, determined atmospheric parameters, and
derived abundances for iron, light, $\alpha$, iron-peak, and neutron-capture
elements. Comparisons to extant measurements reveal general agreement. We did
not find evidence of significant differences between dwarfs and giants, with two
exceptions. Sodium is enhanced in giants with respect to MS stars, as expected
from mixing mechanisms. Lithium shows a normal depletion pattern as a function
of evolutionary phase; in particular Li and \teff \, follow a tight relation for
MS stars, at variance with M67, a cluster of similar age and metallicity, which
shows an unusual (and maybe unique) dispersion in A(Li) at each \teff. 

We reassessed the membership status of all our targets using Gaia parallaxes and
proper motions. Combining Gaia photometry with distance, metallicity, \teff \
and $\log g$ values from our spectroscopic analysis, and \parsec stellar
models,  for Rup 147 we derived distance, Galactic coordinates, reddening, and
age. Ruprecht~147 has metallicity Z=0.017 ([Fe/H]$\approx$0.08 dex), reddening
$E(G_{BP}-G_{RP})=0.15$ (i.e. $E(B-V)=0.113$), an age of 2.5 to 3 Gyr, a
distance about 308 pc from the Sun and 8.28 kpc from the Galactic centre, and
lies 0.07 kpc below the Galactic plane.

With the present paper we add another object to the short list of clusters for
which both giants and dwarfs are analysed, an important test of consistency
between the analysis of the different kinds of stars and of evolutionary
processes affecting the surface abundances.  However, existing spectroscopy is
limited to the brighter part of the MS. Using Gaia data, \cite{cantat18b} found
many more high-probability candidate members of Rup~147 on the single-star MS
and on the well-separated binary sequence, down to the present limit for precise
Gaia astrometry (G=18). There is still room to improve the understanding of this
important cluster and solidify its standing as a benchmark for stellar evolution
studies.

\newpage

\begin{sidewaystable*}
\setlength{\tabcolsep}{.7mm}
\caption{Information for the six stars observed with HARPS-N.}
\begin{tabular}{rcrccrccccccccccccc}
\hline\hline
  CWW   &   Gaia source ID    &    RA        & DEC         & expt  &Date-obs   & UT          &PLX  &ePLX &PMra    &ePMra  &PMde   &ePMde  &G      & $G_{BP}$ & $G_{RP}$    & $RV_{Gaia}$\\ 
        &                          & (J2000)     & (J2000)     & (s)  & (y:m:d)    & (hh:mm:ss)  & (mas)  & (mas)  & (mas/yr)  & (mas/yr) & (mas/yr) & (mas/yr) &    (mag)   &       (mag)  & (mag)  & (km/s)  \\
\hline
 1      &4184143881214538112  &19:15:26.120 &-16:05:57.10 &  900 & 2016-05-24 &01-37-46.061   &3.163 &0.049 &1.709  &0.085 &-25.438 & 0.071 &7.04  & 7.71& 6.30 & 40.56 $\pm$1.89  \\
 2      &4183949198935967232  &19:17:23.840 &-16:04:24.30 &  900 & 2016-05-24 &01:56:40.485   &3.237 &0.049 &-0.537   &0.077 &-26.453 & 0.065 &7.18  & 7.93  & 6.38     & 38.51 $\pm$ 0.17\\
 4      & 4184137077986034048 &19:17:11.300 &-16:03:08.20 & 1200 & 2016-05-24 &02:15:32.621   &3.250 &0.056 &-0.797 &0.089 &-26.997 &0.077 & 8.02 &  8.66& 7.31 & 41.33 $\pm$ 0.18\\
 6      & 4087762027643173248 &19:17:03.430 &-17:03:13.80 & 1200 & 2016-05-24 &02:38:39.041   &3.291 &0.068 &-1.365   &0.108 &-26.523& 0.093 & 8.03 &   8.62    & 7.33    &41.77 $\pm$ 0.19 \\
 10     &4184125807991900928  &19:15:51.290 &-16:17:59.10 & 1200 & 2016-05-24 &03:03:21.454   &3.197 &0.044 &-0.629  &0.076 &-26.895 & 0.071 &8.12 & 8.73&7.42 & 40.80 $\pm$ 0.22\\
 11     &4183930438518525184  &19:18:09.780 &-16:16:22.20 & 1200 & 2016-05-24 &04:44:48.073   &3.368 &0.042 &-1.243  &0.084 &-27.019 & 0.077 &8.25 &8.88&7.53 & 41.62 $\pm$ 0.14 \\
\hline
\end{tabular}
\tablefoot{Observing Programme A33 DDT0. CWW is the identification in \citet{curtis13}. Parallax, proper motions, G, G$_{BP}$,  G$_{RP}$, and RV$_{Gaia}$ come from Gaia DR2 \citep{gaiadr2}. } 
\label{tab1}
\end{sidewaystable*}                                                                                                               

\begin{sidewaystable*}
\setlength{\tabcolsep}{1.2mm}
\caption{Information for the stars in the ESO/HARPS archive.}
\begin{tabular}{rcrrrrrrrrrrrrc}
\hline\hline
  CWW   &   Gaia source ID              &    RA      & DEC         & N. exp  &PLX  &ePLX &PMra    &ePMra  &PMde   &ePMde  &G    & $G_{BP}$ & $G_{RP}$    & $RV_{Gaia}$  \\ 
        &                              & (J2000)     & (J2000)      &        & (mas)  & (mas)  & (mas/yr)  & (mas/yr) & (mas/yr) & (mas/yr) & (mag) &   (mag) &  (mag) & (km/s)\\
\hline
23 &4087853875535923200  & 19:15:42.69 &-16:33:05.0 &15 & 3.388 & 0.070 & -0.707 & 0.157 & -26.717 & 0.104 & 11.38 & 11.71 & 10.91& 42.46 $\pm$1.34  \\
54 &4184136906187904128  & 19:16:55.73 &-16:03:22.0 & 5 & 3.226 & 0.034 & -0.180 & 0.065 & -27.718 & 0.061 & 11.11 & 11.43 & 10.64& 38.64 $\pm$0.73\\
57 &4183934355528796672  & 19:17:04.33 &-16:23:18.5 & 8 & 3.143 & 0.056 & -1.246 & 0.083 & -26.339 & 0.068 & 10.93 & 11.26 & 10.47& 42.84 $\pm$1.88\\
58 &4184188961192147840  & 19:17:21.72 &-15:35:59.2 & 5 & 3.781 & 0.067 &  4.143 & 0.171 & -29.425 & 0.122 & 10.98 & 11.31 & 10.50& 35.58 $\pm$2.06\\
59 &4087799621507741312  & 19:15:12.60 &-17:05:12.1 & 8 & 3.394 & 0.058 & -1.125 & 0.096 & -26.862 & 0.077 & 10.91 & 11.21 & 10.46& 43.73 $\pm$0.84\\
63 &4184157659469838592  & 19:15:29.81 &-15:51:04.7 &13 & 3.174 & 0.062 & -0.553 & 0.086 & -26.958 & 0.068 & 11.20 & 11.55 & 10.71& 40.50 $\pm$1.40\\
70 &4183932534462716928  & 9:16:38.27  &-16:25:03.9 & 2 & 3.169 & 0.052 &  1.136 & 0.081 & -26.436 & 0.070 & 11.19 & 11.51 & 10.74& 48.14 $\pm$1.06 \\
71 &4087860506965490560  & 19:15:45.11 &-16:23:15.7 &19 & 3.196 & 0.044 & -1.350 & 0.085 & -26.558 & 0.062 & 11.64 & 11.99 & 11.15&  \\
74 &4184198788077655936  & 19:15:09.25 &-15:52:24.1 &12 & 3.299 & 0.044 & -0.868 & 0.070 & -27.248 & 0.063 & 11.26 & 11.59 & 10.77& 41.02 $\pm$1.57\\
75 &4183936795070404352  & 19:16:11.21 &-16:21:48.5 &11 & 3.321 & 0.042 & -0.701 & 0.068 & -26.871 & 0.063 & 12.73 & 13.17 & 12.14&39.71 $\pm$4.79 \\
76 &4088004611707768320  & 19:13:43.34 &-16:49:10.9 &11 & 3.194 & 0.095 & -0.696 & 0.108 & -25.737 & 0.078 & 12.27 & 12.66 & 11.73&42.86 $\pm$1.42 \\ 
78 &4184245586042699008  & 19:16:08.79 &-15:24:27.9 &58 & 3.257 & 0.086 & -1.166 & 0.240 & -27.336 & 0.150 & 11.41 & 11.74 & 10.93&\\ 
79 &4088057521393630848  & 19:14:28.16 &-16:20:02.3 &57 & 3.175 & 0.096 & -0.902 & 0.257 & -26.688 & 0.162 & 12.17 & 12.56 & 11.64&43.63 $\pm$1.34 \\ 
81 &4087847067995609728  & 19:15:18.97 &-16:39:24.4 &21 & 3.216 & 0.050 & -2.166 & 0.090 & -25.484 & 0.073 & 11.43 & 11.75 & 10.96&41.59 $\pm$1.33 \\ 
83 &4088110332311492224  & 19:13:41.26 &-16:10:20.1 & 5 & 3.267 & 0.085 & -2.413 & 0.241 & -26.055 & 0.158 & 11.81 & 12.19 & 11.27&\\ 
85 &4087838959097352064  & 19:16:59.40 &-16:35:27.1 &34 & 3.279 & 0.040 & -1.238 & 0.079 & -25.733 & 0.066 & 12.41 & 12.82 & 11.86&42.80 $\pm$2.97 \\ 
90 &4087736159069458304  & 19:16:36.72 &-17:13:10.1 & 8 & 3.252 & 0.038 & -0.618 & 0.069 & -25.935 & 0.062 & 12.09 & 12.45 & 11.57&42.91 $\pm$0.74 \\ 
91 &4184135394358918656  & 19:16:47.25 &-16:04:09.3 & 7 & 3.231 & 0.033 & -0.727 & 0.059 & -26.952 & 0.055 & 12.35 & 12.76 & 11.79&42.20 $\pm$0.68 \\  
93 &4184182737768311296  & 19:16:22.03 &-15:46:15.9 & 6 & 3.085 & 0.040 & -0.937 & 0.065 & -26.009 & 0.057 & 12.55 & 12.98 & 11.98&42.18 $\pm$2.03 \\ 
94 &4184146900561610880  & 19:15:21.41 &-16:00:10.7 &11 & 3.107 & 0.062 & -0.676 & 0.082 & -26.241 & 0.071 & 12.76 & 13.21 & 12.17&42.52 $\pm$0.31 \\ 
97 &4184136558285344000  & 19:17:02.85 &-16:05:16.6 &13 & 3.250 & 0.038 & -0.805 & 0.062 & -27.640 & 0.057 & 12.59 & 13.01 & 12.02&39.82 $\pm$1.17 \\ 
98 &4183940127965076224  & 19:16:26.56 &-16:14:54.5 & 7 & 3.489 & 0.154 & -1.927 & 0.203 & -27.787 & 0.178 & 12.96 & 13.47 & 12.25&44.97 $\pm$2.20 \\ 
\hline
\end{tabular}
\tablefoot{Observing Programmes 091.C-0471, 093.C-0540, and 095.C-0947. CWW is the identification in \citet{curtis13}.
Parallax, proper motions, G, G$_{BP}$,  G$_{RP}$, and RV$_{Gaia}$ come from Gaia DR2 \citep{gaiadr2}. } 
\label{tab2}
\end{sidewaystable*}                                                                                                               

\begin{sidewaystable*}
\setlength{\tabcolsep}{1.25mm}
\caption{Radial velocity, v$\sin~i$, atmospheric parameters, and isochrone-derived stellar mass.}
\begin{tabular}{rrccrccccrccccrrccrcc}
\hline\hline
 CWW  &S/N &RV  & rms  &nr  &\teff &err &log(g) &err &[M/H] &err &$\xi$ &err &N FeI &N FeII&v$\sin~i$ &err  &Instr & mass & err \\  
      & &(km s$^{-1}$) &(km s$^{-1}$)&& (K) &(K) &(dex)&(dex)&(dex)&(dex)&(km s$^{-1}$) &(km s$^{-1}$) & &  &(km s$^{-1}$) &(km s$^{-1}$) & &(M$_\odot$)&(M$_\odot$) \\
\hline  
 1   &40&  37.29 & 0.01 & 1 & 4586 & 34 & 2.47 & 0.08 &  0.20 & 0.14 & 1.55 & 0.01 & 145 & 11 & 3.28 & 0.56 &H-N   & 1.56   &    0.04     \\
 2   &32&  38.68 & 0.01 & 1 & 4373 & 30 & 1.96 & 0.08 &  0.04 & 0.12 & 1.66 & 0.01 & 129 &  9 & 2.94 & 0.62 &H-N   & 1.55   &    0.05     \\
 4   &42&  41.20 & 0.01 & 1 & 4725 & 33 & 2.60 & 0.08 &  0.07 & 0.12 & 1.61 & 0.01 & 148 & 12 & 2.40 & 0.68 &H-N   & 1.52   &    0.04     \\
 6   &47&  41.38 & 0.01 & 1 & 4736 & 33 & 2.51 & 0.08 &  0.07 & 0.11 & 1.51 & 0.01 & 164 & 11 & 2.39 & 0.68 &H-N   & 1.53   &    0.05     \\
10   &50&  40.77 & 0.01 & 1 & 4744 & 33 & 2.38 & 0.08 &  0.07 & 0.11 & 1.50 & 0.01 & 161 & 12 & 2.22 & 0.71 &H-N   & 1.53   &    0.05     \\
11   &47&  41.44 & 0.01 & 1 & 4718 & 32 & 2.60 & 0.08 &  0.15 & 0.10 & 1.48 & 0.01 & 138 & 10 & 2.44 & 0.65 &H-N   & 1.52   &    0.05     \\
23   &115& 41.41 & 0.01 & 15& 6273 & 80 & 4.27 & 0.09 & -0.01 & 0.13 & 1.79 & 0.03 & 214 & 14 & 2.39 & 2.20 & H    & 1.18   &    0.03     \\
71   &114& 41.70 & 0.34 & 19& 6178 & 74 & 4.32 & 0.10 &  0.05 & 0.12 & 1.57 & 0.02 & 218 & 10 & 2.78 & 1.69 & H    & 1.14   &    0.02     \\
75   &50&  42.19 & 0.01 & 11& 5602 & 50 & 4.45 & 0.08 &  0.11 & 0.12 & 1.14 & 0.03 & 246 & 13 & 0.47 & 1.61 & H    & 0.96   &    0.01     \\
76   &57&  42.67 & 0.01 & 11& 5825 & 61 & 4.26 & 0.08 &  0.14 & 0.11 & 1.21 & 0.02 & 235 & 12 &     0.00 & 1.60 & H    & 1.05   &    0.01     \\
78   &182& 41.00 & 0.30 & 59& 6279 & 80 & 4.29 & 0.09 &  0.06 & 0.10 & 1.57 & 0.03 & 212 & 11 & 5.19 & 1.55 & H    & 1.18   &    0.02     \\
79   &150& 41.75 & 0.01 & 58& 5932 & 62 & 4.47 & 0.09 &  0.20 & 0.11 & 1.14 & 0.03 & 246 & 14 &     0.00 & 1.60 & H    & 1.05   &    0.01     \\
81   &148& 41.37 & 0.05 & 21& 6163 & 79 & 4.15 & 0.09 &  0.03 & 0.09 & 1.43 & 0.02 & 240 & 14 &     0.00 & 1.60 & H    & 1.26   &    0.07     \\
83   &49&  41.71 & 0.00 & 5 & 5987 & 77 & 4.43 & 0.10 & -0.04 & 0.13 & 1.42 & 0.03 & 202 &  8 & 2.00 & 1.48 & H    & 1.06   &    0.01     \\
85   &133& 42.66 & 0.02 & 34& 5767 & 62 & 4.42 & 0.09 &  0.07 & 0.11 & 1.27 & 0.02 & 237 & 16 & 2.08 & 1.09 & H    & 1.0    &    0.01     \\
90   &59&  42.75 & 0.04 & 8 & 5994 & 68 & 4.48 & 0.10 &  0.09 & 0.12 & 1.21 & 0.02 & 232 & 13 &     0.00 & 1.60 & H    & 1.07   &    0.01     \\
91   &55&  41.67 & 0.01 & 7 & 5825 & 66 & 4.51 & 0.09 &  0.11 & 0.11 & 1.16 & 0.02 & 220 & 11 &     0.00 & 1.60 & H    & 1.02   &    0.01     \\
93   &43&  41.61 & 0.01 & 6 & 5841 & 57 & 4.53 & 0.09 &  0.18 & 0.13 & 1.19 & 0.03 & 221 &  9 & 1.21 & 1.29 & H    & 1.02   &    0.01     \\
94   &70&  42.06 & 0.01 & 20& 5612 & 49 & 4.49 & 0.09 &  0.15 & 0.14 & 1.15 & 0.03 & 236 & 14 & 1.67 & 0.99 & H    & 0.96   &    0.01     \\
97   &73&  40.36 & 0.02 & 13& 5649 & 53 & 4.29 & 0.09 & -0.04 & 0.11 & 1.27 & 0.03 & 237 & 11 & 1.74 & 1.10 & H    & 0.98   &    0.01     \\
98   &39&  40.68 & 0.04 & 7 & 5380 & 43 & 4.60 & 0.07 & -0.04 & 0.16 & 0.85 & 0.05 & 219 &  7 & 1.78 & 0.89  & H   & 0.9    &    0.01      \\
\hline
58   &67&  36.34 & 2.21 & 9   &&&&&&&&&&&&&  H (bin.)  & & \\
54   &48&  38.84 & 0.02 & 5   &&&&&&&&&&&9.25&2.98 &  H& & \\
57   &87&  41.32 & 0.02 & 8   &&&&&&&&&&&6.78&3.60 &  H& & \\
59   &71&  42.18 & 0.10 & 8   &&&&&&&&&&&12.31&2.73 & H& & \\
63   &92&  41.02 & 0.06 & 13 &&&&&&&&&&&10.06 &2.74 & H& & \\
70   &28&  44.78 & 0.05 & 2   &&&&&&&&&&&14.16&3.23 & H& & \\
74   &96&  41.54 & 0.07 & 12 &&&&&&&&&&&8.12&3.12 & H  & & \\
\hline
\end{tabular}
\tablefoot{For HARPS stars the S/N is given for the co-added spectra. 
For HARPS-N we give the RV and uncertainty on the single RV, 
for HARPS we give the average of the multiple spectra and the rms.
}
\label{param}
\end{sidewaystable*}

\begin{sidewaystable*}
\setlength{\tabcolsep}{1.2mm}
\caption{Light and $\alpha$-element abundances; the Sun is from Grevesse et al. (2007).}
\begin{tabular}{rccccccccccccccccccccc}
\hline
  \multicolumn{1}{c}{CWW} &
  \multicolumn{1}{c}{Li~{\sc i}$_{LTE}$} &
  \multicolumn{1}{c}{upper limit} &
  \multicolumn{1}{c}{Li~{\sc i}$_{NLTE}$} &
  \multicolumn{1}{c}{O~{\sc i}} &
  \multicolumn{1}{c}{Na~{\sc i}$_{LTE}$} &
  \multicolumn{1}{c}{err} &
  \multicolumn{1}{c}{Na~{\sc i}$_{NLTE}$} &
  \multicolumn{1}{c}{err} &
  \multicolumn{1}{c}{Mg~{\sc i}} &
  \multicolumn{1}{c}{err} &
  \multicolumn{1}{c}{Al~{\sc i}} &
  \multicolumn{1}{c}{err} &
  \multicolumn{1}{c}{Si~{\sc i}} &
  \multicolumn{1}{c}{err} &
  \multicolumn{1}{c}{Ca~{\sc i}} &
  \multicolumn{1}{c}{err} &
  \multicolumn{1}{c}{Ti~{\sc i}} &
  \multicolumn{1}{c}{err} &
 \multicolumn{1}{c}{Ti~{\sc ii}} &
  \multicolumn{1}{c}{err} \\
\hline
1   &  0.12  & < &  0.41  &  8.80  &  6.66  &  0.09  &  6.63  &  0.10  &  7.91  &  0.09  &  6.58  &  0.11  &  7.79  &  0.15  &  6.42  &  0.22  &  4.91  &  0.13  &  5.08  &  0.13  \\
2   &  0.29  & < &  0.62  &  8.63  &  6.51  &  0.11  &  6.48  &  0.15  &  7.85  &  0.09  &  6.55  &  0.12  &  7.64  &  0.15  &  6.34  &  0.23  &  4.86  &  0.16  &  4.89  &  0.15  \\
4   &  0.30  & < &  0.54  &  8.77  &  6.50  &  0.08  &  6.47  &  0.12  &  7.82  &  0.10  &  6.50  &  0.15  &  7.68  &  0.14  &  6.37  &  0.20  &  4.88  &  0.15  &  4.97  &  0.13  \\
6   &  0.08  & < &  0.32  &  8.66  &  6.49  &  0.07  &  6.46  &  0.09  &  7.81  &  0.09  &  6.49  &  0.14  &  7.63  &  0.13  &  6.38  &  0.21  &  4.86  &  0.13  &  4.92  &  0.13  \\
10  &  0.32  & < &  0.57  &  8.61  &  6.47  &  0.08  &  6.45  &  0.10  &  7.82  &  0.07  &  6.49  &  0.13  &  7.60  &  0.13  &  6.39  &  0.20  &  4.86  &  0.14  &  4.87  &  0.15  \\
11  &  0.46  & < &  0.71  &  8.74  &  6.55  &  0.07  &  6.56  &  0.13  &  7.90  &  0.10  &  6.59  &  0.12  &  7.72  &  0.13  &  6.46  &  0.20  &  4.94  &  0.14  &  5.03  &  0.14  \\
23  &  2.63  &   &  2.60  &        &  6.33  &  0.10  &  6.26  &  0.13  &  7.66  &  0.16  &  6.31  &  0.09  &  7.48  &  0.12  &  6.36  &  0.12  &  4.88  &  0.19  &  4.92  &  0.08  \\
71  &  2.61  &   &  2.59  &        &  6.33  &  0.12  &  6.29  &  0.13  &  7.65  &  0.17  &  6.40  &  0.05  &  7.52  &  0.13  &  6.41  &  0.14  &  4.88  &  0.13  &  4.96  &  0.08  \\
75  &  0.50  & < &        &        &  6.38  &  0.08  &  6.34  &  0.10  &  7.78  &  0.11  &  6.48  &  0.18  &  7.56  &  0.15  &  6.50  &  0.11  &  5.00  &  0.11  &  4.99  &  0.11  \\
76  &  1.52  & < &  1.57  &        &  6.43  &  0.07  &  6.37  &  0.06  &  7.78  &  0.10  &  6.51  &  0.16  &  7.62  &  0.12  &  6.53  &  0.12  &  4.99  &  0.11  &  5.00  &  0.10  \\
78  &  2.67  &   &  2.65  &        &  6.32  &  0.08  &  6.27  &  0.08  &  7.67  &  0.13  &  6.35  &  0.08  &  7.51  &  0.13  &  6.39  &  0.11  &  4.87  &  0.16  &  4.98  &  0.13  \\
79  &  2.18  &   &  2.21  &        &  6.42  &  0.07  &  6.37  &  0.08  &  7.79  &  0.12  &  6.50  &  0.15  &  7.60  &  0.14  &  6.53  &  0.10  &  5.02  &  0.10  &  5.07  &  0.10  \\
81  &  2.65  &   &  2.63  &        &  6.33  &  0.09  &  6.27  &  0.09  &  7.68  &  0.12  &  6.38  &  0.08  &  7.50  &  0.14  &  6.41  &  0.12  &  4.84  &  0.13  &  4.96  &  0.10  \\
83  &  2.48  &   &  2.47  &        &  6.35  &  0.11  &  6.31  &  0.11  &  7.68  &  0.15  &  6.51  &  0.12  &  7.52  &  0.12  &  6.47  &  0.16  &  4.99  &  0.21  &  4.98  &  0.09  \\
85  &  0.72  & < &        &        &  6.36  &  0.06  &  6.31  &  0.09  &  7.72  &  0.12  &  6.44  &  0.12  &  7.52  &  0.15  &  6.46  &  0.11  &  4.92  &  0.11  &  4.93  &  0.10  \\
90  &  2.32  &   &  2.33  &        &  6.33  &  0.09  &  6.29  &  0.12  &  7.68  &  0.10  &  6.45  &  0.16  &  7.51  &  0.25  &  6.46  &  0.12  &  4.95  &  0.12  &  5.02  &  0.10  \\
91  &  1.78  & < &  1.82  &        &  6.37  &  0.04  &  6.32  &  0.06  &  7.77  &  0.11  &  6.52  &  0.16  &  7.57  &  0.16  &  6.50  &  0.10  &  4.98  &  0.15  &  5.05  &  0.13  \\
93  &  1.59  & < &  1.64  &        &  6.50  &  0.19  &  6.46  &  0.19  &  7.84  &  0.15  &  6.58  &  0.12  &  7.63  &  0.15  &  6.60  &  0.09  &  5.13  &  0.12  &  5.11  &  0.15  \\
94  &  1.06  & < &  1.13  &        &  6.41  &  0.06  &  6.40  &  0.08  &  7.81  &  0.14  &  6.52  &  0.16  &  7.60  &  0.15  &  6.53  &  0.12  &  5.02  &  0.09  &  5.04  &  0.12  \\
97  &  1.26  & < &  1.31  &        &  6.35  &  0.05  &  6.30  &  0.06  &  7.69  &  0.11  &  6.34  &  0.12  &  7.45  &  0.12  &  6.39  &  0.12  &  4.82  &  0.11  &  4.84  &  0.08  \\
98  &  1.01  & < &  1.10  &        &  6.34  &  0.08  &  6.30  &  0.09  &  7.62  &  0.12  &  6.39  &  0.14  &  7.52  &  0.14  &  6.45  &  0.13  &  4.93  &  0.12  &  4.94  &  0.23  \\
\hline
Sun  & 1.05    &  & &     8.66      & 6.17 &         & &      & 7.53       &   & 6.37 &      &   7.51 &      & 6.31 &       & 4.90 &          & 4.90           &   \\
\hline
\end{tabular}
\label{light}
\end{sidewaystable*}

\begin{table*}
\centering
\caption{Iron peak element abundances; the Sun is from Grevesse et al. (2007).}
\begin{tabular}{rccccccccccc}
\hline
\multicolumn{1}{c}{CWW} &
\multicolumn{1}{c}{Sc1}  &
\multicolumn{1}{c}{err}  &
\multicolumn{1}{c}{V1}  &
\multicolumn{1}{c}{err}  &
\multicolumn{1}{c}{Co1}  &
\multicolumn{1}{c}{err}  &
\multicolumn{1}{c}{Ni1}  &
\multicolumn{1}{c}{err}  &
\multicolumn{1}{c}{Cu1}  &
\multicolumn{1}{c}{err}  &
\multicolumn{1}{c}{Zn1}  \\
\hline
1   &  3.08  &  0.06  &  3.95  &  0.14  &  4.98  &  0.14  &  6.35  &  0.22  &  4.25  &  0.12  &  4.31  \\
2   &  3.01  &  0.06  &  3.88  &  0.17  &  4.87  &  0.17  &  6.20  &  0.25  &  4.06  &  0.08  &  4.07  \\
4   &  3.10  &  0.07  &  3.94  &  0.13  &  4.92  &  0.14  &  6.24  &  0.20  &  4.20  &  0.07  &  4.21  \\
6   &  3.05  &  0.07  &  3.89  &  0.12  &  4.87  &  0.14  &  6.22  &  0.20  &  4.15  &  0.06  &  4.24  \\
10  &  3.03  &  0.09  &  3.89  &  0.12  &  4.85  &  0.14  &  6.21  &  0.19  &  4.12  &  0.06  &  4.20  \\
11  &  3.13  &  0.06  &  3.98  &  0.13  &  4.96  &  0.15  &  6.32  &  0.20  &  4.26  &  0.11  &  4.30  \\
23  &  3.36  &  0.20  &  3.92  &  0.21  &  4.89  &  0.25  &  6.18  &  0.16  &  4.22  &  0.07  &  4.30  \\
71  &  3.30  &  0.24  &  3.93  &  0.18  &  4.91  &  0.18  &  6.22  &  0.16  &  4.21  &  0.07  &  4.36  \\
75  &  3.28  &  0.25  &  4.05  &  0.10  &  4.97  &  0.11  &  6.30  &  0.16  &  4.29  &  0.01  &  4.47  \\
76  &  3.28  &  0.23  &  4.02  &  0.14  &  5.00  &  0.12  &  6.32  &  0.15  &  4.29  &  0.04  &  4.51  \\
78  &  3.28  &  0.17  &  3.91  &  0.20  &  4.93  &  0.21  &  6.19  &  0.20  &  4.22  &  0.08  &  4.32  \\
79  &  3.34  &  0.21  &  4.06  &  0.10  &  5.03  &  0.13  &  6.35  &  0.16  &  4.34  &  0.03  &  4.49  \\
81  &  3.32  &  0.24  &  3.87  &  0.19  &  4.84  &  0.18  &  6.20  &  0.14  &  4.18  &  0.06  &  4.40  \\
83  &  3.55  &  0.18  &  4.10  &  0.20  &  4.90  &  0.34  &  6.23  &  0.17  &  4.20  &  0.08  &  4.29  \\
85  &  3.29  &  0.21  &  3.97  &  0.08  &  4.91  &  0.15  &  6.23  &  0.15  &  4.24  &  0.01  &  4.45  \\
90  &  3.27  &  0.20  &  3.97  &  0.15  &  4.98  &  0.16  &  6.29  &  0.16  &  4.29  &  0.09  &  4.48  \\
91  &  3.37  &  0.27  &  4.06  &  0.15  &  5.03  &  0.15  &  6.32  &  0.16  &  4.28  &  0.07  &  4.47  \\
93  &  3.46  &  0.25  &  4.21  &  0.10  &  5.15  &  0.17  &  6.40  &  0.18  &  4.40  &  0.10  &  4.55  \\
94  &  3.31  &  0.17  &  4.09  &  0.09  &  5.01  &  0.12  &  6.32  &  0.16  &  4.24  &  0.08  &  4.49  \\
97  &  3.20  &  0.20  &  3.88  &  0.13  &  4.84  &  0.12  &  6.15  &  0.15  &  4.13  &  0.04  &  4.39  \\
98  &  3.20  &  0.13  &  3.98  &  0.13  &  4.90  &  0.15  &  6.25  &  0.18  &  4.26  &  0.10  &  4.28  \\
\hline
Sun &  3.17  &        &  4.00  &        &  4.92  &            &  6.23  &            &  4.21  &         & 4.60    \\
\hline\end{tabular}
\tablefoot{Cu~{\sc i} abundances corrected, see text.}
\label{heavy}
\end{table*}

\begin{table*}
\centering
\setlength{\tabcolsep}{1.2mm}
\caption{Neutron-capture element abundances; the Sun is from Grevesse et al. (2007).}
\begin{tabular}{rcccccccccccc}
\hline
  \multicolumn{1}{c}{CWW} &
  \multicolumn{1}{c}{Y2}  &
\multicolumn{1}{c}{err}  &
\multicolumn{1}{c}{Zr1}  &
\multicolumn{1}{c}{Zr2}  &
\multicolumn{1}{c}{err}  &
\multicolumn{1}{c}{Ba2}  &
\multicolumn{1}{c}{err}  &
\multicolumn{1}{c}{La2}  &
\multicolumn{1}{c}{err}  &
\multicolumn{1}{c}{Ce2}  &
\multicolumn{1}{c}{err}  &
\multicolumn{1}{c}{Eu2}  \\
\hline
1   &  2.25  &  0.27  &  2.62  &  2.68  &  0.08  &  2.49  &  0.24  &  1.26  &  0.10  &  1.92  &  0.23  &  0.79  \\
2   &  2.08  &  0.28  &  2.55  &  2.41  &  0.08  &  2.19  &  0.37  &  1.04  &  0.18  &  1.69  &  0.13  &  0.66  \\
4   &  2.13  &  0.27  &  2.64  &  2.68  &  0.05  &  2.22  &  0.24  &  1.17  &  0.07  &  1.83  &  0.23  &  0.67  \\
6   &  2.09  &  0.27  &  2.60  &  2.60  &  0.04  &  2.24  &  0.24  &  1.10  &  0.07  &  1.75  &  0.23  &  0.65  \\
10  &  2.01  &  0.20  &  2.57  &  2.54  &  0.01  &  2.21  &  0.24  &  1.02  &  0.05  &  1.66  &  0.23  &  0.58  \\
11  &  2.22  &  0.32  &  2.67  &  2.67  &  0.01  &  2.31  &  0.24  &  1.14  &  0.09  &  1.85  &  0.25  &  0.75  \\
23  &  2.08  &  0.20  &  3.22  &  2.70  &  0.13  &  2.01  &  0.13  &  1.21  &  0.06  &  1.79  &  0.37  &  0.71  \\
71  &  2.16  &  0.22  &  3.13  &  2.78  &  0.21  &  2.04  &  0.21  &  1.14  &  0.04  &  1.71  &  0.42  &  0.71  \\
75  &  2.22  &  0.19  &  3.04  &  2.85  &  0.16  &  2.22  &  0.11  &  1.26  &  0.06  &  2.08  &  0.26  &  0.76  \\
76  &  2.22  &  0.16  &  3.13  &  2.78  &  0.18  &  2.22  &  0.13  &  1.18  &  0.01  &  1.91  &  0.33  &  0.69  \\
78  &  2.13  &  0.19  &  2.97  &  2.67  &  0.14  &  2.10  &  0.18  &  1.21  &  0.10  &  1.77  &  0.42  &  0.63  \\
79  &  2.27  &  0.24  &  2.97  &  2.82  &  0.10  &  2.31  &  0.13  &  1.31  &  0.06  &  2.04  &  0.25  &  0.84  \\
81  &  2.11  &  0.19  &  2.82  &  2.67  &  0.13  &  2.18  &  0.13  &  1.08  &  0.01  &  1.92  &  0.13  &  0.63  \\
83  &  2.20  &  0.27  &  3.45  &  2.79  &  0.08  &  2.16  &  0.14  &  1.09  &  0.08  &  2.12  &  0.06  &  1.01  \\
85  &  2.28  &  0.14  &  3.10  &  2.83  &  0.09  &  2.30  &  0.15  &  1.05  &  0.13  &  2.02  &  0.28  &  0.69  \\
90  &  2.21  &  0.26  &  3.02  &  2.72  &  0.15  &  2.17  &  0.23  &  1.24  &  0.08  &  1.96  &  0.31  &  0.56  \\
91  &  2.30  &  0.33  &  2.74  &  2.74  &  0.16  &  2.21  &  0.06  &  1.39  &  0.11  &  2.02  &  0.37  &  0.84  \\
93  &  2.36  &  0.29  &  3.44  &  2.88  &  0.17  &  2.34  &  0.15  &  1.53  &  0.15  &  2.17  &  0.35  &  0.96  \\
94  &  2.25  &  0.27  &  2.78  &  2.72  &  0.21  &  2.21  &  0.15  &  1.22  &  0.02  &  2.03  &  0.45  &  0.63  \\
97  &  2.27  &  0.15  &  2.89  &  2.88  &  0.11  &  2.28  &  0.13  &  1.22  &  0.05  &  1.92  &  0.33  &  0.64  \\
98  &  2.20  &  0.25  &  2.67  &  2.39  &  0.34  &  2.16  &  0.19  &  0.86  &  0.24  &  2.28  &  0.18  &  0.75  \\
\hline
Sun      &  2.21  &       &  2.58   &  2.58  &        &  2.17  &        &  1.13  &         & 1.70  &        & 0.52    \\
\hline\end{tabular}
\tablefoot{Ba~{\sc ii} and Eu~{\sc ii} abundances corrected, see text.}
\label{ncap}
\end{table*}

\begin{table*}
\centering
\caption{Sensitivity to errors in atmospheric parameters. }
\begin{tabular}{ccccccc}
 \hline
 \multirow{2}{*}{element} &
 \multirow{2}{*}{lines} &
 \multicolumn{2}{c|}{$\Delta$ A(X) (CWW 10)}   &
 \multicolumn{3}{|c}{$\Delta$ A(X) (CWW 81)} \\
 \cline{3-7}
   &  & $\Delta$\teff=+33\,K & $\Delta$log(g)=+0.08\,dex  & $\Delta$\teff=+79\,K & $\Delta$log(g)=+0.09\,dex & $\Delta$[M/H]=+0.06\,dex  \\
 \hline
Li~{\sc i}  & 1  &        &          & 0.070 & 0.000  & 0.010  \\
O~{\sc i}   & 1  &  0.010 &  0.040   &       &        &       \\
Na~{\sc i}  & 8  &  0.029 & -0.005   & 0.035 & -0.005 & -0.006 \\
Mg~{\sc i}  & 5  &  0.018 & -0.010   & 0.028 & -0.012 & -0.004 \\
Al~{\sc i}  & 3  &  0.033 &  0.003   & 0.030 &  0.000 & 0.000 \\
Si~{\sc i}  & 29 & -0.012 &  0.009   & 0.019 &  0.002 & -0.003 \\
Ca~{\sc i}  & 25 &  0.041 & -0.008   & 0.046 & -0.012 & -0.007 \\
Sc~{\sc i}  & 3  &  0.050 &  0.007   & 0.070 & -0.003 & -0.003 \\
Sc~{\sc ii} & 17 &  0.001 &  0.027   & 0.003 &  0.039 & 0.012 \\
Ti~{\sc i}  & 75 &  0.052 &  0.004   & 0.066 & -0.004 & -0.006 \\
Ti~{\sc ii} & 17 & -0.002 &  0.027   & -0.001&  0.036 & 0.014 \\
V~{\sc i}   & 28 &  0.057 &  0.006   & 0.074 &  0.002 & -0.013 \\
Co~{\sc i}  & 28 &  0.020 &  0.014   & 0.061 & -0.001 & -0.004 \\
Ni~{\sc i}  & 84 &  0.013 &  0.011   & 0.047 & -0.001 & -0.004 \\
Cu~{\sc i}  &  3 &  0.023 &  0.020   & 0.053 &  0.003 & -0.010 \\
Zn~{\sc i}  &  1 & -0.010 &  0.010   & 0.030 &  0.010 & -0.010 \\
Y~{\sc ii}  & 13 &  0.012 &  0.025   & 0.013 &  0.028 & 0.015 \\
Zr~{\sc i}  &  1 &  0.060 &  0.000   & 0.080 &  0.000 & -0.010 \\
Zr~{\sc ii} &  2 &  0.010 &  0.040   & 0.010 &  0.040 & 0.005 \\
Ba~{\sc ii} &  3 &  0.013 &  0.010   & 0.033 &  0.020 & 0.017 \\
La~{\sc ii} &  3 &  0.008 &  0.033   & 0.015 &  0.035 & 0.015 \\
Ce~{\sc ii} &  3 &  0.003 &  0.033   & 0.005 &  0.035 & 0.015 \\
Eu~{\sc ii} &  1 &  0.000 &  0.030   & 0.010 &  0.040 & 0.020 \\
 \hline
 \end{tabular}
 \tablefoot{Sensitivity computed for star CWW~10, 
a giant star with \teff= 4744$\pm$33\,K, log(g)= 2.38$\pm$0.08\,dex, [M/H]=-0.02$\pm$0.06\,dex;
and a dwarf star CWW 81, with \teff=6163$\pm$79\,K, log(g)=4.15$\pm$0.09\,dex, [M/H]=+0.02$\pm$0.06\,dex.}
\label{sens}
\end{table*}

\begin{table*}
\caption{Average abundance ratios for giants, dwarfs, and entire sample.}
\begin{tabular}{l ccr ccr ccr}
\hline
[X/Fe] & mean &rms &num &mean &rms &num &mean &rms &num \\
\hline
&\multicolumn{3}{c}{giants} &\multicolumn{3}{c}{dwarfs}&\multicolumn{3}{c}{all} \\
\cmidrule(lr){2-4} \cmidrule(lr){5-7} \cmidrule(lr){8-10} 
Fe  & 0.10 &0.06 &6 & 0.07 &0.08 &15 & 0.08 &0.07 &21 \\
O1  &-0.06 &0.05 &6 &      &     &   &-0.06 &0.04 & 6 \\
Na1 & 0.24 &0.02 &6 & 0.08 &0.05 &15 & 0.13 &0.08 &21 \\
Mg1 & 0.22 &0.03 &6 & 0.12 &0.04 &15 & 0.15 &0.06 &21 \\
Al1 & 0.06 &0.05 &6 & 0.01 &0.04 &15 & 0.02 &0.06 &21 \\
Si1 & 0.07 &0.03 &6 &-0.04 &0.04 &15 &-0.01 &0.06 &21 \\
Ca1 &-0.02 &0.03 &6 & 0.09 &0.05 &15 & 0.06 &0.07 &21 \\
Sc1 &-0.20 &0.05 &6 & 0.08 &0.11 &15 & 0.00 &0.16 &21 \\
Sc2 &-0.09 &0.04 &6 &-0.05 &0.05 &15 &-0.06 &0.05 &21 \\
Ti1 &-0.11 &0.03 &6 &-0.02 &0.06 &15 &-0.05 &0.07 &21 \\
Ti2 &-0.04 &0.03 &6 & 0.02 &0.04 &15 & 0.00 &0.05 &21 \\
V1  &-0.18 &0.04 &6 &-0.07 &0.08 &15 &-0.10 &0.08 &21 \\
Co1 &-0.11 &0.02 &6 &-0.04 &0.04 &15 &-0.06 &0.05 &21 \\
Ni1 &-0.07 &0.01 &6 &-0.04 &0.04 &15 &-0.05 &0.04 &21 \\
Cu1 &-0.14 &0.04 &6 &-0.03 &0.05 &15 &-0.06 &0.07 &21\\
Zn1 &-0.48 &0.04 &6 &-0.25 &0.04 &15 &-0.32 &0.11 &21\\
Y2  &-0.18 &0.04 &6 &-0.06 &0.07 &15 &-0.10 &0.08 &21 \\
Zr1 &-0.07 &0.05 &6 & 0.37 &0.23 &15 & 0.25 &0.28 &21\\
Zr2 &-0.08 &0.07 &6 & 0.10 &0.11 &15 & 0.05 &0.13 &21\\
Ba2 & 0.01 &0.05 &6 &-0.05 &0.08 &15 &-0.03 &0.08 &21 \\
La2 &-0.11 &0.05 &6 & 0.00 &0.11 &15 & 0.04 &0.11 &21\\
Ce2 &-0.02 &0.05 &6 & 0.21 &0.16 &15 & 0.15 &0.17 &21\\
Eu2 & 0.06 &0.03 &6 & 0.15 &0.14 &15 & 0.12 &0.12 &21 \\
\hline
\end{tabular}
\tablefoot{All abundances are [X/Fe], with the exception of [Fe/H]. Li is not reported here. The value for Na is in NLTE. 
}
\label{mean}
\end{table*}

\begin{table*}
\caption{Comparison with literature for RV and atmospheric parameters. }
\begin{tabular}{rccccccccccccc}
\hline\hline
CWW &RV & \teff & $\log g$ & [Fe/H] & $\xi$ &RV$_{C14}$
     &RV$_C$ &RV$_G$ & \teff & $\log g$ & [Fe/H] & $\xi$ & Ref \\
     \hline
&\multicolumn{5}{c}{Present paper} & &\multicolumn{6}{c}{Literature} & \\
\cmidrule(lr){2-6} \cmidrule(lr){7-13} 
  1 &37.29 &&&& &38.0 & 38.5  &$40.56\pm1.89$& & & & & 2,3,5\\
  2 &38.68 &&&&  &39.0 & 43.4  &$38.51\pm0.17$& & & & & 2,3,5  \\
   4 &41.20 &&&& &         & 42.7  &$41.33\pm0.18$& & & & & 2,5\\
  6 &41.38 &&&&  &42.1 & 46.2   &$41.77\pm0.19$& & & & & 2,3,5\\
10 &40.77 &4744 &2.38 &0.07 &1.50 &       & 40.1  &$40.80\pm0.22$&4633 &2.53 & 0.14 &1.28 & 1,5\\
11 &41.44 &&&&  &41.8 & 44.2  &$41.62\pm0.14$    & & & & & 2,3,5\\
23 &41.41    &&&&&&&$42.46\pm1.34$ &&&&&5\\
75&42.19    &&&&&&&$39.71\pm4.79$ &&&&&5\\
76&42.67    &&&&&&&$42.86\pm1.42$ &&&&&5\\
78 &41.00 &6279 &4.29 &0.06 &1.57 &     &41.02  &&6129 &3.60 &-0.01 & & 2\\
79&41.75    &&&&&&&$43.63\pm1.34$ &&&&&5\\
81&41.37    &&&&&&&$41.59\pm1.33$ &&&&&5\\
85&42.66    &&&&&&&$42.80\pm2.97$ &&&&&5\\
90&42.75    &&&&&&&$42.91\pm0.74$ &&&&&5\\
91 &41.67 &5825 &4.51 &0.09 &1.16 &     &40.35  &$42.20\pm0.68$ &5747 &4.35 & 0.06 & & 2,5\\
93 &41.61 &5841 &4.53 &0.09 &1.19 &     &41.58 &$42.18\pm2.03$ &5697 &4.453 & 0.141 & & 4,5\\
94&42.06    &&&&&&&$42.52\pm0.31$ &&&&&5\\
97&40.36   &&&&&&&$39.82\pm1.17$ &&&&&5\\
98&40.68    &&&&&&&$44.97\pm2.20$ &&&&&5\\
\hline
\end{tabular}
\tablefoot{
(1) \cite{pakhomov} ; (2) \cite{curtis13}, Tables 3, 5 ; (3) \cite{carlberg14} ; (4) \cite{curtis18}; (5) Gaia DR2} 
\label{conf}
\end{table*}

\newpage

\begin{acknowledgements}
This paper is based on observations made with the Italian Telescopio Nazionale Galileo (TNG) operated on the island of La Palma by the Fundaci\'ion Galileo Galilei of the INAF (Istituto Nazionale di Astrofisica) at the Spanish Observatorio del Roque de los Muchachos of the Instituto de Astrofisica de Canarias. This paper is based on data obtained from the ESO Science Archive Facility under request number 299084.
XF acknowledges funding by Premiale 2015 MITiC (PI B. Garilli) and the EU COST Action CA16117 (ChETEC). 
This work presents results from the European Space Agency (ESA) space mission Gaia. Gaia data are being processed by the Gaia Data Processing and Analysis Consortium (DPAC). Funding for the DPAC is provided by national institutions, in particular the institutions participating in the Gaia MultiLateral Agreement (MLA). The Gaia mission website is https://www.cosmos.esa.int/gaia. The Gaia archive website is https://archives.esac.esa.int/gaia.
This research has made use of Vizier and SIMBAD, operated at CDS, Strasbourg,
France,  NASA's Astrophysical Data System, and {\sc TOPCAT} (http://www.starlink.ac.uk/topcat/, \citealt{topcat}).  This research made use of the cross-match service provided by CDS, Strasbourg.
\end{acknowledgements}

\end{document}